\ifx\pdfoutput\undefined        
  \documentclass[a4paper]{aa}
  \usepackage{url}
\else                           
  \documentclass[pdftex,a4paper]{aa}
  \usepackage[bookmarks=false]{hyperref}
\fi
\usepackage{graphicx}
\newcommand{\EQ}{\begin{equation}}
\newcommand{\EN}{\end{equation}}
\newcommand{\EQA}{\begin{eqnarray}}
\newcommand{\ENA}{\end{eqnarray}}

\newcommand{\Fig}[1]{Fig.~\ref{#1}}

\newcommand{\meanuu}{\overline{\mbox{\boldmath $u$}}{}}{}
{}
{}
{}
{}
{}
\newcommand{\rrr}{\mbox{\boldmath $r$} {}}
\newcommand{\xx}{{\vec{x}}}

\newcommand{\uu}{{\vec{u}}}

\newcommand{\FF}{{\vec{F}}}

\newcommand{\grav}{\mbox{\boldmath $g$} {}}
\newcommand{\nab}{\vec{\nabla}}
\newcommand{\OO}{\mbox{\boldmath $\Omega$} {}}

\newcommand{\SSSS}{\mbox{\boldmath ${\sf S}$} {}}

\newcommand{\DD}{{\rm D} {}}

\newcommand{\dd}{{\rm d} {}}
\newcommand{\const}{{\rm const}  {}}

\def\half{{\textstyle{1\over2}}}

\def\onethird{{\textstyle{1\over3}}}

\newcommand{\yapj}[3]{ #1, {ApJ,} {#2}, #3}

\newcommand{\yana}[3]{ #1, {A\&A,} {#2}, #3}

\newcommand{\yprl}[3]{ #1, {PRL,} {#2}, #3}
\newcommand{\ypre}[3]{ #1, {PRE,} {#2}, #3}

\newcommand{\ymn}[3]{ #1, {MNRAS,} {#2}, #3}

\newcommand{\yjcp}[3]{ #1, {J. Comput. Phys.,} {#2}, #3}
\newcommand{\yjour}[4]{ #1, {#2}, {#3}, #4}

\begin{document}

\title{The angular momentum transport by the strato-rotational instability in simulated Taylor-Couette flows}
\author{A. Brandenburg\inst{1} \and G. R\"udiger\inst{2}}

\institute{
NORDITA, Blegdamsvej 17, DK-2100 Copenhagen \O, Denmark
\and 
Astrophysikalisches Institut Potsdam, An der Sternwarte 16, D-14482 Potsdam, Germany
}

\date{\today}

\abstract{}
{To investigate the stability and angular momentum transport by the
strato-rotational instability in the nonlinear regime.
}{ 
The hydrodynamic compressible equations are solved in a cartesian box
in which the outer cylinder is embedded.
Gravity along the rotation axis leads to density stratification.
No-slip boundary conditions are used in the radial direction, while free-slip
conditions are used on the two ends of the cylinders.
}{
The strato-rotational instability is confirmed and the Reynolds stress is
shown to transport angular momentum away from the axis.
However, the growth rate decreases with increasing Reynolds number.
This, as well as the presence of boundaries renders this instability
less relevant for astrophysical applications.
}{}
\titlerunning{The angular momentum transport by the strato-rotational instability}
\maketitle

\section{Introduction}

The magneto-rotational instability (MRI) is now commonly invoked in
order to explain the origin of turbulence in accretion discs
(Balbus \& Hawley 1998).
This is because there exists no purely hydrodynamic local instabilities
in accretion discs in the sense that periodic shearing box calculations
do not yield instability.
Already when vertical shear is included, there is a linear instability,
but for thin discs its growth rate is small compared with that of the MRI
(Urpin \& Brandenburg 1998; Arlt \& Urpin 2004).
Purely hydrodynamic instabilities are of interest for protostellar
discs where the conductivity is usually so poor that magnetic effects
are unimportant, and hence the MRI may be irrelevant.
Another possibility for a hydrodynamic instability is the nonlinear
shear instability (Chagelishvili et al.\ 2003,
Afshordi, Mukhopadhyay \& Narayan 2005).
The general difficulty with such models has been discussed by
Balbus, Hawley \& Stone (1996).
However, a conclusive resolution of this problem using simulations
remains difficult because of the large Reynolds numbers required.

In an attempt to address the possibility of hydrodynamic instabilities
in accretion discs, one can resort to experiments.
All these experiments have boundaries, so their relevance to accretion
discs is questionable.
Nevertheless, such experiments remain interesting in their own right.
The possibility of a nonlinear instability of a Taylor-Couette flow
experiments has recently been addressed by Richard \& Zahn (1999).
The issue was revived by recent findings of
Molemaker, McWilliams \& Yavneh (2001) who found the possibility of a
{\it linear} instability in Taylor-Couette flow with density
stratification along the axis.
Dubrulle et al.\ (2005) confirmed this instability for a range of different
radial boundaries conditions.

More recent work by Shalybkov \& R\"udiger (2005), who considered regular
Taylor-Couette flow in the presence of no-slip boundaries, gave a detailed
stability criterion for the instability, which is henceforth referred to
as the strato-rotational instability (SRI). Withjack \& Chen (1974)
observed nonaxisymmetric patterns in density-stratified Taylor-Couette
flows close and  beyond the Rayleigh line.

Umurhan (2005) shows that, within the framework of
the quasi-hydrostatic semi-geostrophic approximation, the instability
only survives in the presence of no-slip radial boundaries.
The purpose of the present paper is to demonstrate the existence of this
instability using direct simulations and to calculate its saturation
level in the nonlinear regime.
In contrast to nonlinear instabilities, the SRI can be studied conveniently
by simulations, because the critical Reynolds numbers are well below $10^3$.
We are thus also able to compute sign and, in the nonlinear regime,
also the amount of angular momentum transport due to the SRI.
Let us emphasize here that the sign of
angular momentum transport (inwards or outwards) is nontrivial.
There are several examples in the literature (mostly for rotating
convection) where the cross correlation
\EQ
Q_{r\phi}=\overline{u'_ru'_\phi}
\label{Q1}
\EN
of the one-point correlation tensor 
\EQ
Q_{ij}=\overline{u'_i(\xx,t)  u'_j(\xx,t)}
\label{Q2}
\EN
turns out to be negative, i.e.\ angular momentum is transported inwards
by the rotating turbulence (see Ryu \& Goodman 1992;
Kley, Papaloizou \& Lin 1993; Stone \& Balbus 1996). 
The  primes in (\ref{Q1}) and (\ref{Q2}) denote the departures from the
toroidally averaged flow, i.e.\ $\uu'=\uu-\meanuu$, where
\EQ
\meanuu={\textstyle\int_0^{2\pi}}\uu\,\dd\phi/2\pi.
\EN
Here we shall find that the zonal flux of angular momentum,  (\ref{Q1}),
always proves to be positive  i.e.\ the angular momentum  is transported
outwards by the SRI.

\section{The model}

We consider the governing equations for a compressible perfect gas,
\EQ
{\DD\ln\rho\over\DD t}=-\nab\cdot\uu,
\EN
\EQ
{\DD\uu\over\DD t}=\grav-c_{\rm s}^2(\nab\ln\rho+\nab s/c_p)+\FF_{\rm visc}
+\FF_{\rm rot}+\FF_{\rm drive},
\label{dudt}
\EN
\EQ
\rho T{\DD s\over\DD t}=\nab\cdot\left(c_p\rho\chi\nab T\right),
\EN
where $\uu$ is the velocity, $\rho$ the density, $s$ the specific entropy,
$\grav=\mbox{const}$ is gravity,
$c_p$ and $c_v$ are the specific heats at constant pressure and volume,
respectively, $\gamma=c_p/c_v=5/3$ is their assumed ratio,
$c_{\rm s}$ is the sound speed, $T$ is the temperature, and the two
are related to the other quantities via
\begin{equation}
(\gamma-1)c_pT=
c_{\rm s}^2=c_{\rm s0}^2\exp\left[\gamma s/c_p+(\gamma-1)\ln\rho/\rho_0\right].
\end{equation}
Here, $c_{\rm s0}$ and $\rho_0$ are normalization constants.
These constants also define the zero point of the specific entropy.
The viscous force is
\EQ
\FF_{\rm visc}=\nu\left(\nabla^2\uu+\onethird\nab\nab\cdot\uu
+2\SSSS\cdot\nab\ln\rho\right),
\EN
where ${\sf S}_{ij}=\half(u_{i,j}+u_{j,i})-\onethird\delta_{ij}
\nab\cdot\uu$ is the traceless rate of strain tensor, where commas
denote partial differentiation.
We consider a rotating frame of reference, such that in that frame
the angular velocity of the outer cylinder is zero.
The resulting apparent force is
\EQ
\FF_{\rm rot}=-2\OO_{\rm out}\times\uu+\Omega_{\rm out}^2\vec{\varpi},
\EN
where $\vec{\varpi}=(x,y,0)$ is the cylindrical radius vector.
In our frame of reference, the angular velocities of the outer
and inner cylinders are imposed by a body force of the form
\EQ
\FF_{\rm drive}=
-\tau^{-1}\left[(\uu-\uu_{\rm in})\xi_{\rm in}(\rrr)
+\uu\xi_{\rm out}(\rrr)\right],
\EN
where $\xi_{\rm in}(\rrr)=1$ when $|\rrr|<R_{\rm in}$ and 0 otherwise,
and $\xi_{\rm out}(\rrr)=1$ when $|\rrr|>R_{\rm out}$ and 0 otherwise,
$\tau$ is a relaxation time (comparable to the time step), and
$\uu_{\rm in}=(\OO_{\rm in}-\OO_{\rm out})\times\rrr$ is the linear
velocity of the inner cylinder in the rotating frame.
The temperature is given by $c_p T=c_{\rm s}^2/(\gamma-1)$.
Gravity is written as $\grav=(0,0,-g)$, where the value of $g$ is varied.

The {\it initial} stratification is isothermal
($c_{\rm s}=c_{\rm s0}=\const$) with $\ln\rho=-z/H$ and $s/c_p=(\gamma-1)z/H$,
where $H=c_{\rm s0}^2/(\gamma g)$ is the density scale height.
The Brunt-V\"ais\"al\"a frequency $N$ is given by
\EQ
N^2={g\over c_p}{\dd s\over \dd z}=(\gamma-1){g^2\over c_{\rm s}^2}
=\left(1-{1\over\gamma}\right){g\over H},
\EN
which, following Shalybkov \& R\"udiger (2005),
is expressed in terms of the Froude number,
$\mbox{Fr}=\Omega_{\rm in}/N$.
For $\gamma=5/3$ and $g=1$ we have $N=0.27\,\Omega_{\rm in}$, so
$\mbox{Fr}=3.7$.

Viscosity and thermal diffusivity are specified in terms of
Reynolds and Prandtl numbers,
\EQ
\mbox{Re}=\Omega_{\rm in}R_{\rm in}(R_{\rm out}-R_{\rm in})/\nu,
\quad
\mbox{Pr}=\nu/\chi,
\EN
respectively.
In all cases considered below we assume $\mbox{Pr}=10$, i.e.\ thermal
diffusion is small compared with viscosity.
(We need to keep a certain minimum amount of thermal diffusivity
to stabilize the code.)

We assume the radial boundaries to be insulating, i.e.\
$\partial T/\partial\varpi=0$,
and for the velocity we adopt a no-slip boundary condition, i.e.\
\EQ
\uu=\uu_{\rm in}\quad\mbox{(on $|\rrr|=R_{\rm in}$)},\quad
\uu=0\quad\mbox{(on $|\rrr|=R_{\rm out})$}.
\EN
In the vertical direction we adopt insulating free-slip boundary conditions.
The vertical extent of the cylindrical shell is in all cases $L_z=4$.
It is convenient to normalize the various quantities with respect to their
values on the inner cylinder and to introduce the ratios
\EQ
\hat\eta\equiv R_{\rm in}/R_{\rm out},\quad
\hat\mu\equiv \Omega_{\rm out}/\Omega_{\rm in}.
\EN
Following Shalybkov \& R\"udiger (2005), the flow is
strato-rotationally unstable when
\EQ
\hat\eta^2<\hat\mu<\hat\eta
\quad\mbox{(SRI, nonaxisymmetric)}.
\EN
For $\hat\mu<\hat\eta^2$ the flow is always Rayleigh unstable
both to axisymmetric and to nonaxisymmetric perturbations.
In the following we consider the cases
$\hat\eta=0.5$ and 0.25 ($R_{\rm out}=2$ and 4, using $R_{\rm in}=1$),
so we expect the SRI to occur in the ranges $0.25<\hat\mu<0.5$
and $0.06<\hat\mu<0.25$ in the two cases.
For $\hat\mu<0.25$ (or 0.06) the flow is Rayleigh-unstable.

For the runs presented below the maximum flow speed is at the most 1.2
in units where $\Omega_{\rm in}=R_{\rm in}=1$.
In order that the flow remains everywhere subsonic,
i.e.\ $\max(|\uu|)<c_{\rm s}$, we choose $c_{\rm s0}=3$,
which means that the sound speed varies around 3.
For the SRI compressibility is however not essential.

We model the problem by embedding the cylinder into a box where the
velocities corresponding to rigid inner and outer cylinders is prescribed
(see Dobler, Shukurov \& Brandenburg 2002 for details regarding this approach).
We solve the equations in cartesian geometry using the Pencil Code\footnote{
\url{http://www.nordita.dk/software/pencil-code}}, which is a
memory-efficient sixth-order finite difference code using the
$2N$-RK3 scheme of Williamson (1980).
For most of the cases we consider a Reynolds number of 500 where
a resolution of $128^3$ meshpoints is sufficient.
In a few cases we considered larger Reynolds numbers where
a higher resolution of up to $512^3$ meshpoints was used.

\section{Results}

\subsection{Dependence on $\hat\mu$}

\begin{figure}[t!] 
\centering\includegraphics[width=\columnwidth]{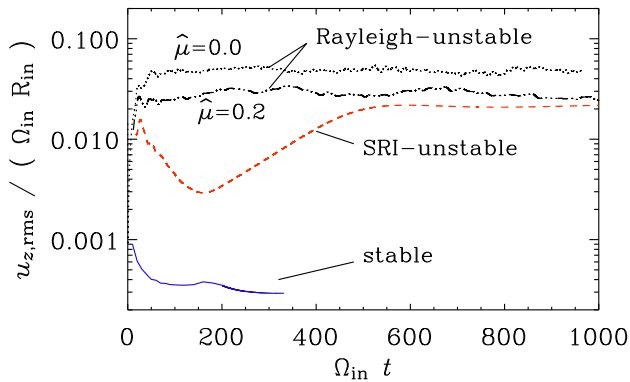}\caption{
Amplitudes for $\hat\mu=0$, 0.2, 0.3, and 0.6.
Here, $g/(R_{\rm in}\Omega_{\rm in}^2)=1$ and $\mbox{Re}=500$.
}\label{uzrms_comp}\end{figure}

As a simple diagnostics of the instability we use the root mean square
velocity, $u_{z,\rm rms}$.
In \Fig{uzrms_comp} we plot the evolution of $u_{z,\rm rms}$ for different
values of $\hat\mu$.
As expected, no instability is found for $\hat\mu=0.6$.
For smaller values ($\hat\mu=0$ and 0.2), the flow is Rayleigh unstable.
For $\hat\mu=0.3$ the flow is strato-rotationally unstable (\Fig{uzrms_comp}).
Since such flows are stable to axisymmetric perturbations by the
Solberg-H{\o}iland criteria (R\"udiger, Arlt \& Shalybkov 2002),
the instability must be necessarily nonaxisymmetric, which is
indeed the case\footnote{
Animations of the flow, showing also the nonaxisymmetric
nature of the instability, can be found on
\url{http://www.nordita.dk/~brandenb/movies/couette/}}.

\subsection{Dependence on $g$}

\begin{figure}[t!]
\centering\includegraphics[width=\columnwidth]{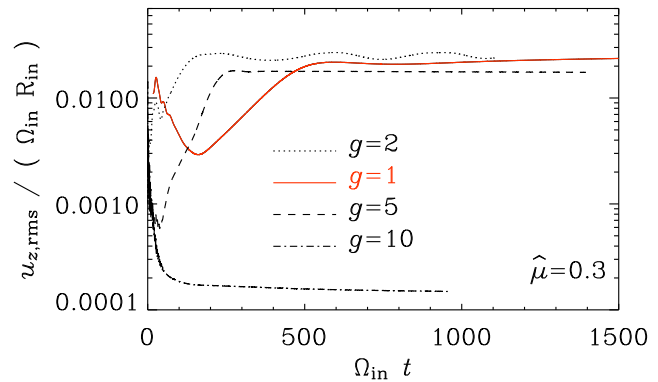}\caption{
Amplitudes for $g/(R_{\rm in}\Omega_{\rm in}^2)=1$, 2, 5, and 10,
using $\hat\mu=0.3$ and $\mbox{Re}=500$.
}\label{uzrms_compg}\end{figure}

\begin{figure}[t!] 
\centering\includegraphics[width=\columnwidth]{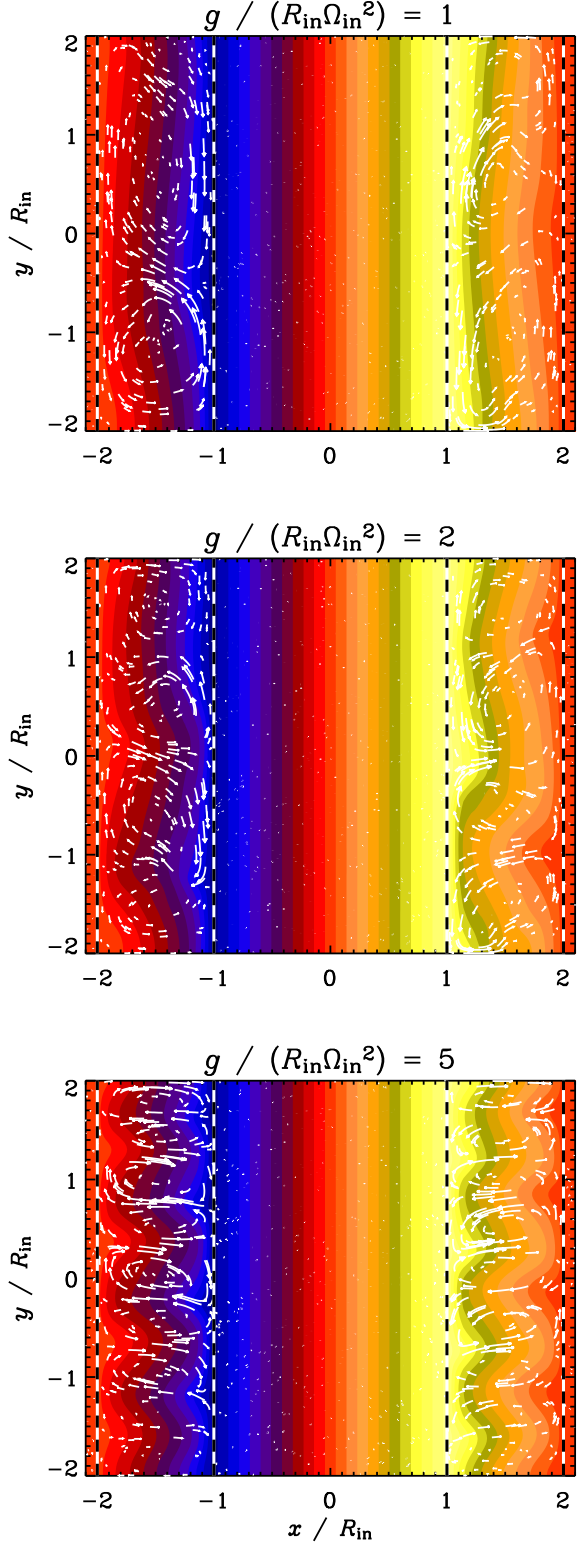}\caption{
Velocity pattern in the meridional plane for different
values of $g/(R_{\rm in}\Omega_{\rm in}^2)$ and $\mbox{Re}=500$.
}\label{puxz_all}\end{figure}

For weak stratification (small values of $g$)
the growth rate of the instability is small.
For small gap width, $\mbox{Re}=500$ and
$g/R_{\rm in}\Omega_{\rm in}^2=2$ we find the largest rms velocity.
For $g/R_{\rm in}\Omega_{\rm in}^2=5$ the growth rate is maximum while
for $g=10$ no instability is found; see \Fig{uzrms_compg}.
However, when Re is increased to 1000, also the case with $g=10$
yields instability.
The number of azimuthal cells increases as the value of $g$ is increased;
see \Fig{puxz_all}, where we show the velocity field in a meridional
cross-section.
Note that for $g=5$ there are 4 eddies in the $z$ direction.
Since $L_z/R_{\rm in}=4$, the vertical wavelength is 1.

\subsection{Critical Reynolds number}

\begin{figure}[t!]
\centering\includegraphics[width=\columnwidth]{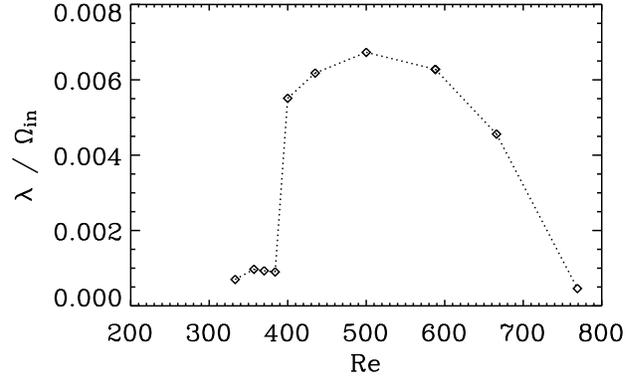}\caption{
Dependence of the growth rate $\lambda$ on the Reynolds number $\mbox{Re}$.
The critical value is about $\mbox{Re}\approx400$.
For $\mbox{Re}>500$, $\lambda$ decreases with increasing $\mbox{Re}$.
}\label{lam-re}\end{figure}

The growth rates are on the order of $(0.01...0.02)\times\Omega_{\rm in}$,
which is relatively small.
The dependence of the growth rate $\lambda$ on the Reynolds number $\mbox{Re}$
is shown in \Fig{lam-re}.
For $\mbox{Re}<400$ the vertical rms velocity seems to grow in an
oscillatory fashion.
However, for $\mbox{Re}>500$, the growth rate decreases with increasing
Reynolds number.

\subsection{Reynolds stress}

In order to estimate the ability of the flow to transport angular
momentum, we compute the radial component of the Reynolds stress.
It turns out that $Q_{r\phi}$ is positive throughout most of the
domain, so the flow is transporting angular momentum outwards.
The normalized stress, $Q_{r\phi}/(R_{\rm in}^2\Omega_{\rm in}^2)$
is around $5\times10^{-4}$ for $g/(R_{\rm in}\Omega_{\rm in}^2)$
around 2 and 5.
For smaller values of $g$, the vertical scale of the eddies becomes
comparable to or exceeds the vertical extent of the cylinders, so the
SRI begins to be suppressed.
Likewise, for larger values of $g$, the vertical scale of the eddies
becomes comparable to the scale where viscosity effects can still be
regarded as weak.

The fact that the ratio $Q_{r\phi}/(R_{\rm in}^2\Omega_{\rm in}^2)$
is much less than unity suggests that
angular momentum transport is relatively inefficient.
This supplements the earlier impression based on the small value of
the normalized growth rate of the instability.

\begin{table}[htb]\caption{
Dependence of the Reynolds stress on gravity  for container with small gap.
}\vspace{12pt}\centerline{\begin{tabular}{cccc}
\hline
$g/(R_{\rm in}\Omega_{\rm in}^2)$ &
$Q_{r\phi}/(10^{-4}R_{\rm in}^2\Omega_{\rm in}^2)$ &
$Q_{zz}/(10^{-4}R_{\rm in}^2\Omega_{\rm in}^2)$ \\
\hline
0.5&$>0.16$& $>0.04$ \\ 
 1 & 0.9 & 4.6 \\ 
 2 & 4.3 & 5.4 \\ 
 5 & 4.2 & 1.1 \\ 
 \hline
\label{Tsum2}\end{tabular}}\end{table}

\begin{table}[htb]\caption{
Dependence of the Reynolds stress on gravity for the container with large gap.
Note that for $g/(R_{\rm in}\Omega_{\rm in}^2)=2$ the perturbations
are slowly decaying.
}\vspace{12pt}\centerline{\begin{tabular}{cccc}
\hline
$g/(R_{\rm in}\Omega_{\rm in}^2)$ &
$Q_{r\phi}/(10^{-4}R_{\rm in}^2\Omega_{\rm in}^2)$ &
$Q_{zz}/(10^{-4}R_{\rm in}^2\Omega_{\rm in}^2)$ \\
\hline
 2 & 0.1  & 0.02 \\ 
 5 & 3    & 0.6 \\ 
 \hline
\label{Tsum3}\end{tabular}}\end{table}

\begin{figure}[t!] 
\centering\includegraphics[width=\columnwidth]{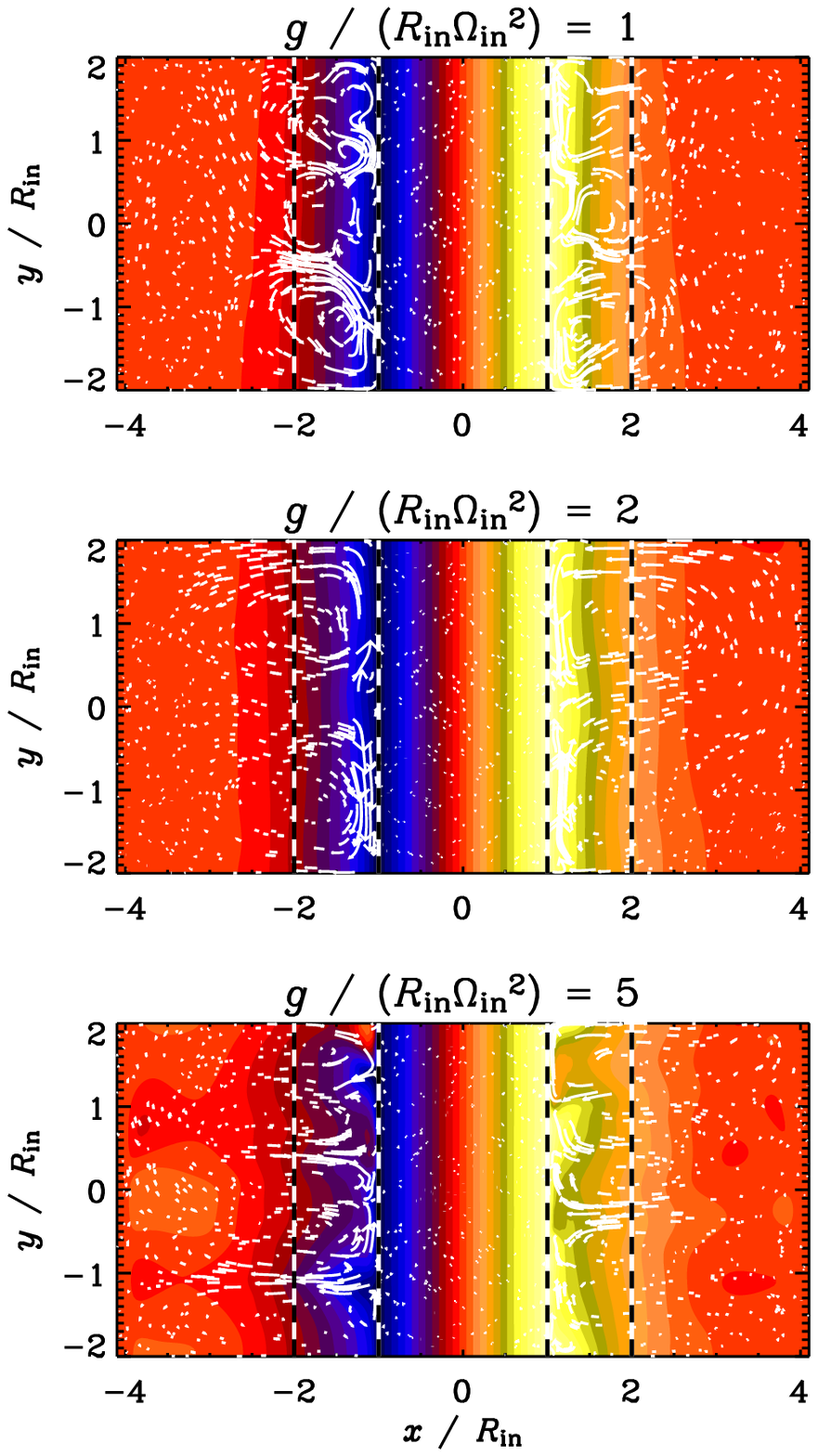}\caption{
Velocity pattern in the meridional plane for different
values of $g/(R_{\rm in}\Omega_{\rm in}^2)$.
(For $g/(R_{\rm in}\Omega_{\rm in}^2)\le2$ the perturbations
are slowly decaying.)
}\label{puxz_wide}\end{figure}

Let us now express the Reynolds stress in astrophysically relevant terms
by normalizing it with respect to $H_{\rm in}^2\Omega_{\rm in}^2$, where
we have chosen the inner radius as reference.
(We note that in accretion discs, $H_{\rm in}\Omega_{\rm in}$ is
proportional to the sound speed.)
The Reynolds stress normalized in this way is referred to as the
Shakura-Sunyaev alpha,
\EQ
\alpha_{\rm SS}=Q_{r\phi}/\left(H_{\rm in}^2\Omega_{\rm in}^2\right).
\label{alfass2}
\EN
Using the data from Table~\ref{Tsum2}, we find
\EQ
\alpha_{\rm SS}=(1...5)\times10^{-4}\left(H_{\rm in}/R_{\rm in}\right)^{-2}.
\label{alfass}
\EN
One may be tempted to insert here the standard result for
Shakura-Sunyaev discs, $H/R=0.03$, which would suggest rather large
values of the stress normalized in this way; $\alpha_{\rm SS}=0.1...0.5$.
However, this argument assumes that the instability survives for
small values of $H/R$, which is unfortunately not true.
Furthermore, for larger radial the instability seems to develop
only near the inner cylinder; see \Fig{puxz_wide}.
Nevertheless, provided the flow is unstable, the resulting normalized
stress, $Q_{r\phi}/(R_{\rm in}^2\Omega_{\rm in}^2)$, is still roughly
the same.

\section{Conclusions}

In the present work we have explored the nonlinear saturation of the
strato-rotational instability using a finite difference method.
We verify the stability criterion of the Taylor-Couette flow
in the presence of vertical stratification.
In agreement with earlier results by Shalybkov \& R\"udiger (2005),
a nonaxisymmetric instability is found when the azimuthal velocity
on the inner cylinder, $R_{\rm in}\Omega_{\rm in}$, exceeds that on
the outer cylinder, $R_{\rm out}\Omega_{\rm out}$.
This leads to instability in a regime that would be stable by the
Solberg-H{\o}iland criteria, which only apply to axisymmetric flows on in
the tight-winding approximation (R\"udiger, Arlt \& Shalybkov 2002).

Increasing $\mbox{Re}$ sufficiently beyond the critical value for the
instability, the growth rate decreases with increasing $\mbox{Re}$.
This is related to the fact that the flow is driven via the viscous force,
which is obviously not relevant in astrophysics, but it is relevant
to the experiments that are currently being designed to address these
questions that are astrophysically motivated in other ways.

\begin{acknowledgements}
We thank the organizers and participants
of the program ``Magnetohydrodynamics of Stellar
Interiors'' at the Isaac Newton Institute in Cambridge (UK) for
a stimulating environment that led to the present work.
The Danish Center for Scientific Computing is acknowledged
for granting time on the Linux cluster in Odense (Horseshoe).
\end{acknowledgements}

\end{document}